# Molecular Dynamics Simulation for the Analysis of Mechanical Properties and Effect of Stone-Wales and Bi-Vacancy Defect on Carbon Nanotube Reinforced Iron Composites

Raashiq Ishraaq[1,a)], Santosh Chhetri[1,b)], Omkar Gautam[1,c)], Shahriar Muhammad Nahid[2], A. M. Afsar[1]

[1]Department of Mechanical Engineering, Bangladesh University of Engineering and Technology, Dhaka-1000, Bangladesh
[2]Department of Mechanical Science and Engineering, University of Illinois Urbana Champaign, Urbana, Illinois 61801, United States

Corresponding author: a)ishraaqraashiq@gmail.com ,b)santosh1410181@gmail.com ,c)omkargautam500@gmail.com

**Abstract.** Carbon nanotube (CNT) reinforced metal matrix composites (MMCs) are gaining the attention of the researchers because of their demand in space and automobile industries for having low weight and high mechanical properties. Iron is the most used metal in all engineering fields. Therefore, reinforcing iron with CNT can reduce its required amount, which might have a positive economic impact due to the reduced cost of production. However, before the industrial application of any material the mechanical properties under different conditions must be known. In this study, the mechanical properties of iron reinforced separately with single, double and triple wall CNTs are investigated by Molecular Dynamics (MD) simulation. The study revealed that the strength and stiffness of pure iron could be enhanced up to 80.4 % and 57.4 %, respectively, by adding CNTs into iron. We also investigated the effect of fiber volume percentage and temperature on the mechanical properties of the composite having single, double and triple-walled carbon nanotubes individually. As the stone-wales and bi-vacancy defects are inherently introduced in CNTs during manufacturing, their effect on mechanical properties are also investigated in the present study.

## INTRODUCTION

CNT discovered by Sumlo Iijima created a revolution in many scientific fields due to their outstanding thermal, electronic and mechanical properties [1]. Because of their high stiffness (∼1TPa) and strength (∼ 63GPa), they are considered as a good reinforcing agent in materials. However, most of the previous studies used CNT as reinforcement for polymeric composites. Now researchers are looking into metal matrices, due to the widespread application of metals [2].

MD simulation can be used to predict the influence of mechanical reinforcement at a microscopic level for composites. Failure mechanism and the influence of various parameters on the strength can be also calculated, which can provide the trend of the mechanical properties in the actual situation. Faria et al. [3] explored the tensile and compressive behavior of copper-CNT (Cu-CNT) composite. Their study showed that CNT enhances the Young's modulus of copper by 32% and tensile strength by 14%. Silvestre et al. [4] investigated the enhancement of mechanical properties for adding CNT in aluminum by applying compressive load on single-crystal aluminum-CNT composite (Al-CNT) and found that the Young's modulus increased up to 100 %. Both of these studies were done performing MD.

Iron can be considered a good matrix material due to its widespread use. Study relating to iron matrices is very scarce. Recently Parswajinan et al. [5] studied the mechanical properties of multiwall carbon nanotube (MWCNT) reinforced iron composite (Fe-MWCNT), which was created by powder metallurgy technique. Goyal et al. [6] synthesized iron-CNT (Fe-CNT) composite by chemical vapor disposition and found CNT increases the yield strength of pure iron by 45%.

In spite of iron being the most used metal, no MD studies were performed on iron composites to study and predict their behavior. Motivated by this fact in this study we conducted a series of MD simulations of uniaxial tension on iron composites reinforced by CNT of different numbers of walls for measuring the possible increase in strength and stiffness compared to pure Fe. We also investigated how the mechanical properties of the composite vary with fiber volume percentage and temperature for containing single and multiwall CNT. Bi-vacancy and Stone-Wales defect drastically reduce the strength of CNT. These defects occur in different layers of multiwall carbon nanotubes (MWCNT) depending on the manufacturing process. In order to choose an optimum manufacturing process, we must know how the presence of these defects in different layers of the CNT affects the mechanical properties of the composite. Here, we also studied how the presence of these defects in different walls of double-walled CNT affects the mechanical parameters of the composite.

## METHODOLOGY

In our study, we constructed our atomistic model of representative volume element (RVE) as a rectangular prism of iron with square surface area, which was embedded with CNT having different number of walls (Fig 1).Visual molecular dynamics (VMD) software was used to create the CNTs and Large-scale Atomic/Molecular Massively Parallel Simulator (LAMMPS) was used to create the iron matrix. The CNTs layers in the z direction. The specimen's length in the z direction was kept constant at 54.264 Å. The side length of the square surface area was changed to vary the matrix volume percentage. By changing the percentage of matrix volume the fiber volume percentage was also changed. The side length was changed to 62.832 Å, 57.12 Å, 51.408 Å, and 45.696 Å to make the fiber volume 11 %, 14 %, 17 %, and 20 % respectively. Periodic boundary condition (PBC) was employed in all directions to simulate the conditions inside the bulk composite. Figure 1 shows the detailed overview of the modeled RVE having 14% fiber volume, containing single-walled carbon nanotube (SWCNT), double-walled carbon nanotube (DWCNT) and triple-walled carbon nanotube (TWCNT) separately. While modeling the MWCNTs, the relationship between the chirality of an outer and the next inner CNT was maintained by following the $m=n+5$ rule [7]. Where $m$ is the chirality of outer CNT and $n$ is the chirality of inner CNT. It was done to ensure the graphitic distance of 3.4 Å between two consecutive CNT walls [8]. The outermost CNT of all the RVE models has a chirality of (14, 14). LAMMPS software package was also used to study the tensile response and The Open Visualization Tool (OVITO) was used as a post-processing software.

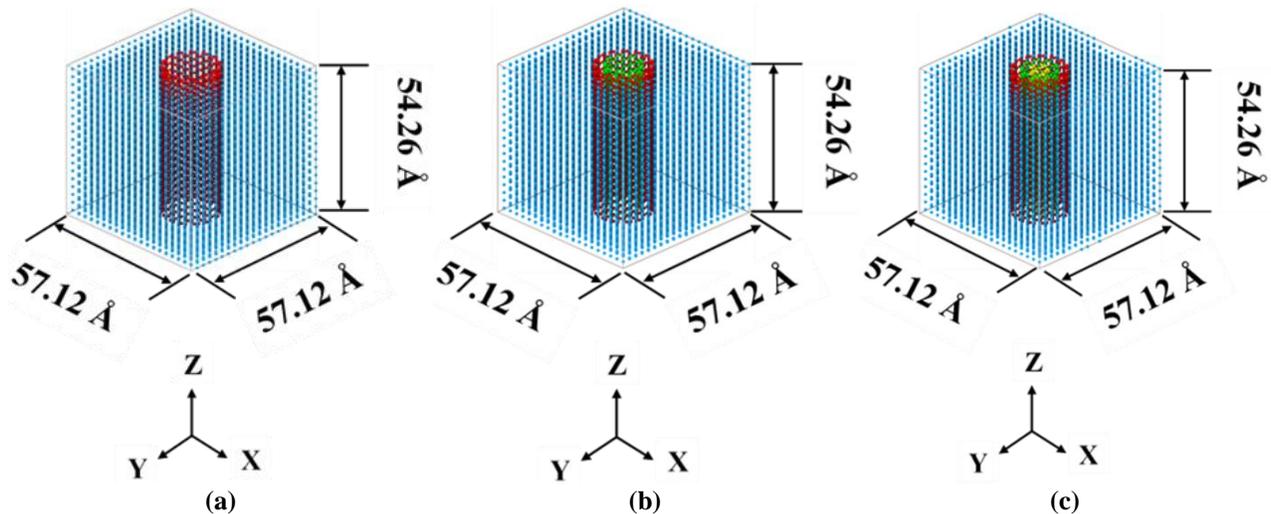

**FIGURE 1.** (a), (b), (c) shows RVE containing triple-walled, double-walled and single-walled CNTs accordingly

Stone-wales and bi-vacancy defects were modeled in the inner and outer wall of the DWCNT separately for investigating how the presence of these defects on different walls affects the mechanical property of the composite. Figure 2 shows the modeled defects.

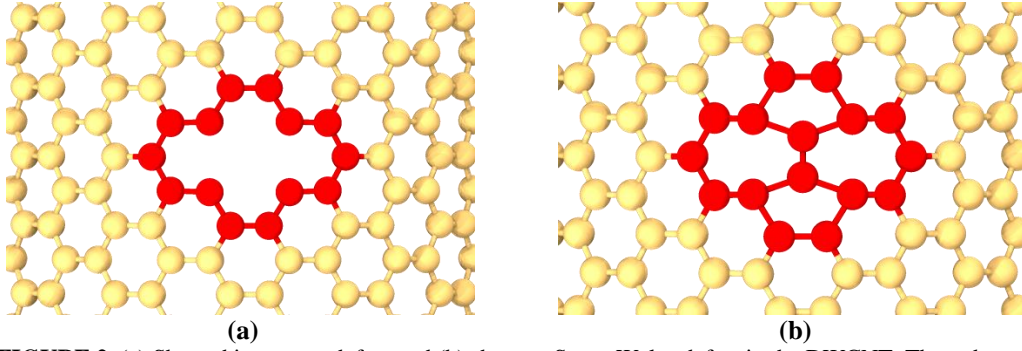
**FIGURE 2.** (a) Shows bi-vacancy defect and (b) shows a Stone-Wales defect in the DWCNT. The red atoms constitute unit cells that are affected by the defects.

Three potentials were used to model the interactions between the atoms. For Fe-Fe interactions, embedded atom method (EAM) potential developed by Ackland et al. [9] was used. Mathematically, EAM potential is described as below:

$$E_i = F_\alpha\left(\sum_{j\neq i} \rho_\beta(r_{ij})\right) + \frac{1}{2}\sum_{j\neq i} \emptyset_{\alpha\beta}(r_{ij}) \qquad (1)$$

Here $F$ is the embedding energy, which is a function of electron density $\rho$. $\emptyset$ is pair potential function and $r_{ij}$ is the distance between atom $i$ and $j$. Subscripts $\alpha$ and $\beta$ are element types of atom $i$ and $j$.

AIREBO potential was used for the C-C bonds in the CNT. It consists of REBO, Lennard jones, and torsional interaction potential as shown:

$$E = \sum_i \sum_{i\neq j}\left[E_{ij}^{REBO} + E_{ij}^{LJ} + \sum_{k\neq i}\sum_{l\neq i,j,k} E_{i,j,k,l}^{TORSION}\right] \qquad (2)$$

Here, $E_{ij}^{REBO}, E_{ij}^{LJ}$ and $E_{i,j,k,l}^{TORSION}$ are Rebo, Lennard-Jones and torsional potential respectively. The REBO potential describes the covalent bonding. The Lennard-Jones and torsional terms describe the intermolecular interaction and dihedral angle effect.

To simulate the side interfacial interaction of Fe and C atoms Vander wall's interaction was considered [10]. 6-12 Lennard-Jones potential was used to define the Fe-C interaction as described below:

$$E = 4\epsilon\left[\left(\frac{\sigma}{r}\right)^{12} - \left(\frac{\sigma}{r}\right)^{6}\right] \qquad (3)$$

$\epsilon$ is the potential well depth, $r$ is the interatomic distance and $\sigma$ is the distance where potential energy is zero. The value $\sigma$ (3.01715 Å) and $\epsilon$ ( 0.03095 eV) used in this simulation were found by using the Lorentz-Berthelot rule [10] and from the parameters used in other studies [11,12]. The radius of the hole inside the matrix was such that the CNT and iron atoms maintain a distance of h=0.8584$\sigma$ as suggested by Jiang et al.[13].

After creating the model, energy minimization using the conjugate gradient method is done to ensure there was no overlap of atoms. A sequence of NVT and NPT equilibration was also performed for 100000 time steps each to ensure the temperature and pressure fluctuation is stable. The value of each time step was 1 fs. During the NPT equilibration, the temperature was held at 300K and pressure at 1 bar to simulate atmospheric conditions. The RVE was stretched along the z-axis with a constant strain rate of 0.001 ps$^{-1}$ to simulate the tensile stress.

The mechanical stress was calculated using the virial stress as described below:

$$\sigma = \frac{1}{\Omega}\sum_i\left[-m_i \dot{u}_i \otimes \dot{u}_i + \frac{1}{2}\sum_{i\neq j} r_{ij} \otimes f_{ij}\right] \qquad (4)$$

Here $m$ is the mass of the atom $i$ and $\dot{u}_i$ is the time derivative of the displacement vector $u_i$. $f_{ij}$ is the interatomic force on atom $i$ by atom $j$ and $r_{ij}$ is the distance between them.

## VALIDATION

For validating our simulation process we performed MD simulation of uniaxial tension on an RVE of pure iron and SWCNT of (10, 10) chirality to compare with other studies. The strength of the CNT is 105 GPa which matches closely with the experiments performed by Eatemadi et al [14]. During the simulation of iron, the strain rate was changed to 0.01 ps$^{-1}$ for maintaining similar conditions with the study done by Mayer [15]. The size of the modeled RVE of pure iron was same as the literature. The yield strength of iron was found around 11.3 GPa which matches closely with the results found by Mayer[15].

Since there are no previous MD studies of Fe-CNT, we validated our process with the experimental results of Goyal et al. [6]. For that, we constructed another RVE model with CNT of chirality (6, 6) having 2.2% fiber volume and applied uniaxial tension with a strain rate of $0.001ps^{-1}$. Uniaxial tension was also applied to pure iron of similar size and with similar conditions. Yield strength was obtained by using 0.4% strain offset technique to maintain similarity with the experimental test. The yield strength of composite and pure iron obtained from this simulation is 7.5 GPa and 5.35 GPa respectively, which is higher than the results found in the experiment. This discrepancy between the values of yield strength obtained from simulation and experiment can be explained by the fact that real materials contain many defects and the strain rate in MD simulations is always higher due to computational limitations. In spite of that, the percentage of increase in strength and stiffness of pure iron for adding CNT is 40%. Which is very similar to the result observed in the experiment ($\approx$ 45%). As a result, the simulation accurately predicts the trend of mechanical parameters for adding CNT in Fe. A similar method for validation was used by Cong et al [16] to validate their simulation with the experiment.

## RESULT AND DISCUSSION

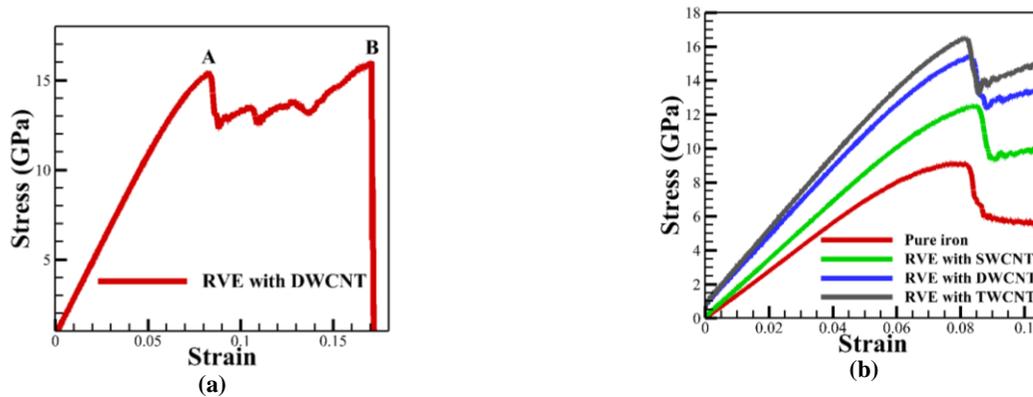

**FIGURE 3.** Stress-strain diagram of (a) DWCNT reinforced iron composite up to the failure of the CNT. (b) pure iron and composites reinforced with different types of CNTS.

Figure 3-a shows the stress-strain curve of the RVE model containing DWCNT (Fig 1-b) and having 14% fiber volume. All other stress-strain curves of the RVE models show a similar pattern. At first, the matrix fails near a strain of 0.08, which is the first drop in stress (point A). However, the composite is not re-usable after point A though it continues to bear the load. After point A, the stress rises and falls numerous times due to strain hardening. The CNT breaks near 0.17 strain (Point B) and then the composite fails to bear any load. Figure 3-b shows the stress-strain relation of pure iron and Fe-CNT having 14% fiber volume and reinforced by SWCNT, DWCNT and TWCNT separately. Stress-strain graphs were plotted until the failure of the matrix because the composite was unusable after it. The stress developed when the matrix fails was also considered the ultimate strength. The strength of the composite increased 36.6%, 68.97%, 80.4%, and stiffness increased 21.12%, 46.25%, 57.4% than pure iron for containing single, double and triple wall carbon nanotube respectively. From Fig 3-b it is obvious that the level of enhancement of strength and stiffness is quite low for adding a third wall in the CNT compared to the level of enhancement for adding the second wall. This phenomenon is observed because having a third wall slightly reduces the amount of void present inside the CNT. So the fiber volume fraction increases comparatively less for adding a third wall. Therefore, according to the mixture rule, the strength and stiffness increase at a lower rate.

# Effect of Fiber Volume Percentage on Single and Multiwall CNT Reinforced Fe Composite

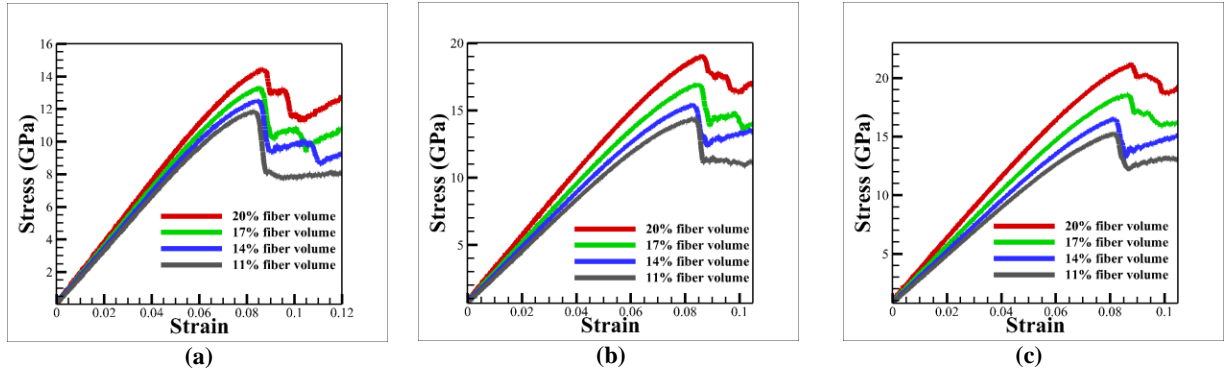

(a) (b) (c)

**FIGURE 4.** This figure shows the stress-strain relation of the composite for different fiber volume percentages. (a), (b) and (c) is the stress-strain diagram of RVE containing SWCNT, DWCNT, and TWCNT respectively.

Figure 4 shows stress-strain graphs of the composite reinforced separately by SWCNT, DWCNT, and TWCNT for various fiber volume percentage. For every 3% increase in fiber volume the Young's modulus of the composite reinforced by SWCNT, DWCNT and TWCNT increased on an average of 4.9%, 8.3%, and 9.5% respectively than the previous state. The ultimate strength also increased on an average of 6.7 %, 9.9% and 11.58% for composites reinforced by SWCNT, DWCNT, and TWCNT respectively. So the mechanical properties of the composite reinforced with CNT having a higher number of walls vary the most with fiber volume percentage. Therefore, it can be concluded, when more layers of CNT are present the mechanical parameters become much more sensitive to the percentage of fiber volume.

# Effect of the Presence of Stone-Wales and Bi-Vacancy Defect in Different Walls of DWCNT

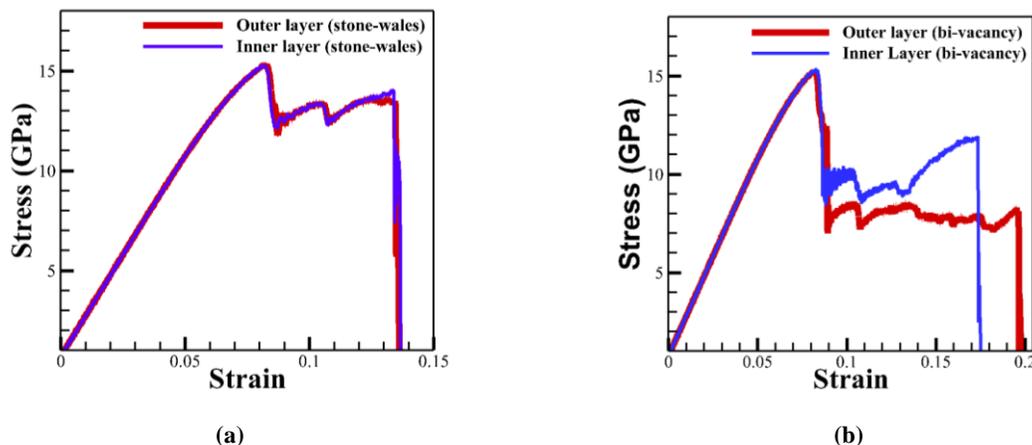

(a) (b)

**FIGURE 5:** The figure shows the stress-strain relationship of RVE containing defected DWCNT and having 14% fiber volume. The stress-strain relationship curves are shown up to the failure of the CNT. (a) Show the stress-strain relations when stone-defect is present on the outer (red) and the inner (blue) layer of the DWCNT. Similarly, (b) shows the stress-strain relation for vacancy defect.

Figure 5-a shows two stress-strain relations of two composites having 14% fiber volume and reinforced with DWCNT but has a Stone-Wales defect in a different wall of the DWCNT. From the graphs, it can be seen that the stress-strain relationship is similar up to the failure of the matrix (≈ 0.08 strain) but a slight deviation can be seen at the stress when CNT fails which is quite negligible.

However, in the case of the bi-vacancy defect (Fig 5-b), the stress-strain relation changes drastically depending on which layer of DWCNT the defect is present. The stress-strain relation remains the same up to the failure of the matrix (≈ 0.08 strain) in both cases. After the failure of the matrix the composite having a bi-vacancy defect in the outer wall of the DWCNT continues to develop lower stress than the composite having defect in the inner wall of the DWCNT. However, the failure strain is higher in the composite that has defect in the outer wall of the DWCNT than the composite having defect in the inner wall. From Fig 5-b, it can also be observed that the area under the stress-strain curve (which is also the amount of strain energy) after the failure of the matrix is equal for both cases. Therefore, it can be concluded that the composite stores strain-energy at a higher rate if bi-vacancy defect is present in inner wall of the DWCNT than if it is present in the outer wall.

## Effect of Temperature on Single and Multiwall CNT Reinforced Iron Composite

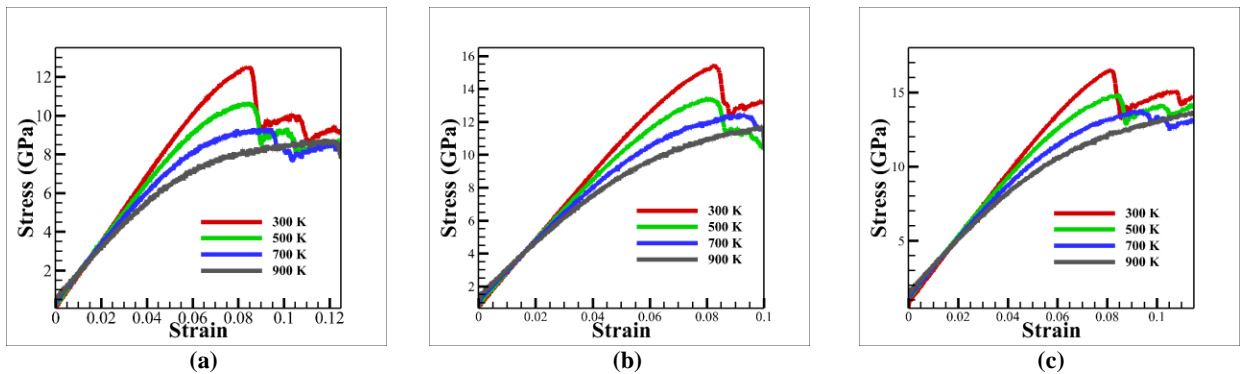

**FIGURE 6.** (a), (b) and (c) respectively show the stress-strain relationship of the composite separately reinforced by SWCNT, DWCNT, and TWCNT at different temperatures.

Uniaxial tension simulation was performed at various temperatures on RVE models having 14% fiber volume, which are separately reinforced by SWCNT, DWCNT, and TWCNT. Figure 6 shows the stress-strain relation obtained from the simulation. These graphs are plotted until the failure of the matrix since the composite is not re-usable after that. From these graphs, it can be seen that strength and stiffness fall drastically for all the RVE models but plasticity increases near the failure point as the temperatures get high. For every 200K increase in temperatures the ultimate strength of the composite containing SWCNT, DWCNT and TWCNT decrease on an average of 12.25 %, 8.68 %, and 8.67 % respectively compared to the composite at previous temperature. From Fig 6 it is noticed that at high temperatures the stress doesn't decrease much after the failure of the matrix. This indicates that the percentage of the load carried by the matrix decreases as the temperature gets high. Incrementing the temperature every 200K also decreases the stiffness of the composite compared to the previous temperature by an average of 6.8 %, 6.2 % and 5.9 % for composite reinforced by SWCNT, DWCNT, and TWCNT respectively. Therefore, the variation of stiffness with temperature is less for RVE containing CNT having a higher number of walls. This phenomenon occurs because the stiffness of CNT varies less with temperature than iron and having CNT of higher walls increases the mass percentage of CNT in the composite. Therefore, the overall property of the composite becomes aligned with the property of CNT and so its stiffness varies less with temperature.

## CONCLUSIONS

The mechanical properties of carbon-nanotube reinforced iron were investigated in this paper by molecular dynamics. The results obtained from the simulation revealed that the strength and stiffness of pure iron increased significantly depending on the number of layers of CNT. Our investigation also found that mechanical properties become more sensitive to fiber volume percentage for composites containing CNT having a higher number of walls. From our analysis on defects, it can also be concluded that bi-vacancy defect drastically changes the stress-strain relation depending on which layer the defect is present but stone-wales defect does not show this phenomenon. Furthermore, temperature analysis revealed that the mechanical properties of composites containing CNT having more walls varies less with temperature. In addition, the model of RVE used in this study can be further used to investigate the effect of fiber orientation, fiber coating and defect in the matrix material.

## REFERENCES

1. J.N. Coleman, U. Khan, W.J. Blau, Y.K. Gun'ko, Small but strong: A review of the mechanical properties of carbon nanotube-polymer composites, Carbon N. Y. 44 (2006) 1624–1652. doi:10.1016/j.carbon.2006.02.038.
2. S.R. Bakshi, D. Lahiri, A. Agarwal, Carbon nanotube reinforced metal matrix composites - a review, Int. Mater. Rev. 55 (2010) 41–64. doi:10.1179/095066009X12572530170543.
3. B. Faria, C. Guarda, N. Silvestre, J.N.C. Lopes, D. Galhofo, Strength and failure mechanisms of cnt-reinforced copper nanocomposite, Compos. Part B Eng. 145 (2018) 108–120. doi:10.1016/j.compositesb.2018.02.033.
4. N. Silvestre, B. Faria, J.N. Canongia Lopes, Compressive behavior of CNT-reinforced aluminum composites using molecular dynamics, Compos. Sci. Technol. 90 (2014) 16–24. doi:10.1016/j.compscitech.2013.09.027.
5. C. Parswajinan, B. Vijaya Ramnath, C. Elanchezhian, S. V. Pragadeesh, P.R. Ramkishore, V. Sabarish, Investigation on mechanical properties of nano ferrous composite, Procedia Eng. 97 (2014) 513–521. doi:10.1016/j.proeng.2014.12.276.
6. A. Goyal, D.A. Wiegand, F.J. Owens, Z. Iqbal, Enhanced yield strength in iron nanocomposite with in situ grown single-wall carbon nanotubes, J. Mater. Res. 21 (2006) 522–528. doi:10.1557/jmr.2006.0061.
7. A. Lopez-bezanilla, W. Zhou, J. Idrobo, Electronic and Quantum Transport Properties of Atomically, (2014).
8. V. Harik, Mechanics of Multiwall Carbon Nanotubes, Mech. Carbon Nanotub. (2018) 165–189. doi:10.1016/b978-0-12-811071-3.00007-x.
9. G.J. Ackland, D.Y. Sun, M. Asta, M.I. Mendelev, S. Han, D.J. Srolovitz, Development of new interatomic potentials appropriate for crystalline and liquid iron, Philos. Mag. 83 (2003) 3977–3994. doi:10.1080/14786430310001613264.
10. F. Banhart, Interactions between metals and carbon nanotubes: At the interface between old and new materials, Nanoscale. 1 (2009) 201–213. doi:10.1039/b9nr00127a.
11. P. The, T. The, E. Use, O. Basic, S. For, E. Health, E. Isbn, Simulatian of self diffusion of iron ( Fe ) and Chromium ( Cr ) in Liquid lead by Molecular Dynamic, (n.d.) 207–208.
12. A. Mokhalingam, D. Kumar, A. Srivastava, Mechanical Behaviour of Graphene Reinforced Aluminum Nano composites, Mater. Today Proc. 4 (2017) 3952–3958. doi:10.1016/j.matpr.2017.02.295.
13. L.Y. Jiang, Y. Huang, H. Jiang, G. Ravichandran, H. Gao, K.C. Hwang, B. Liu, A cohesive law for carbon nanotube/polymer interfaces based on the van der Waals force, J. Mech. Phys. Solids. 54 (2006) 2436–2452. doi:10.1016/j.jmps.2006.04.009.
14. H. Eatemadi, Ali Daraee, H. Karimkhanloo, M. Kouhi, N. Zarghami, A. Akbarzadeh, M. Abasi, Y. Hanifehpour, S. Joo, Carbon nanotubes: properties, synthesis, purification, and medical applications, Nanoscale Res. Lett. 9 (2014) 393. doi:10.1186/1556-276X-9-393.
15. A.E. Mayer, Dynamic shear and tensile strength of iron: Continual and atomistic simulation, Mech. Solids. 49 (2014) 649–656. doi:10.3103/s0025654414060065.
16. Z. Cong, S. Lee, Study of mechanical behavior of BNNT-reinforced aluminum composites using molecular dynamics simulations, Compos. Struct. 194 (2018) 80–86. doi:10.1016/j.compstruct.2018.03.103.